\documentclass[runningheads]{llncs}
\usepackage{listings}
\usepackage{graphicx}
\usepackage{booktabs}
\usepackage{fancyvrb}
\usepackage{todonotes}
\usepackage{xcolor} 
\newcommand{\nbval}{\texttt{nbval}{}}
\newcommand{\Nbval}{\texttt{Nbval}{}} 
\newcommand{\pytest}{\texttt{pytest}{}}
\newcommand{\Pytest}{\texttt{Pytest}{}} 
\begin{document}

\title{Testing with Jupyter notebooks: \\ {NoteBook} {VALidation} (\nbval) \\
  plug-in for \pytest}
%
\titlerunning{Testing with Jupyter notebooks: \nbval}
%
\author{Hans Fangohr \inst{1,2} \and
Vidar Fauske\inst{3} \and
Thomas Kluyver\inst{1, 2} \and
Maximilian Albert \inst{2} \and
Oliver Laslett \inst{2} \and
David Cortes \inst{2} \and
Marijan Beg\inst{2} \and
Min Ragan-Kelly\inst{3}}
\authorrunning{H. Fangohr et al.}
%
\institute{European XFEL GmbH, Holzkoppel 4, 22869 Schenefeld, Germany\\
 \and
University of Southampton, Southampton SO17 1BJ, United Kingdom \\
\and
Simula Research Laboratory, Martin Linges vei 25, 1364 Fornebu, Norway\\
}
\maketitle              
\begin{abstract}
  The Notebook validation tool \nbval{} allows to load and execute Python
  code from a Jupyter notebook file. While computing outputs from the cells
  in the notebook, these outputs are compared with the outputs saved in the
  notebook file, treating each cell as a test. Deviations are reported as test
  failures, with various configuration options available to control the behaviour.
  Application use cases include the validation of notebook-based documentation, tutorials and
  textbooks, as well as the use of notebooks as additional unit, integration and
  system tests for the libraries that are used in the notebook. \Nbval{} is implemented as a plugin for
  the pytest testing software.

%
%
\end{abstract}

\section{Introduction}
\label{sec:org0a766b0}

Software engineering for research comes with
a number of challenges, including the need for rapid prototyping and
experimentation; performing computational and data science studies;
testing and reproducibility of results; documenting research,
algorithms and software; and creating figures for publications.

Jupyter Notebooks~\cite{Jupyter} can help address many of these
challenges. Historically, it was difficult to verify that a collection of existing
notebooks can be re-executed and execute correctly after some time (as the
underlying libraries may have changed).

Here, we motivate and explain the design and implementation of the
NoteBook VALidation tool \nbval, which helps close the gap with
regard to reproducibility, testing and documentation.

Nbval can be useful during all stages of a project's lifecycle. In
particular, it can help support an iterative workflow where tests and
documentation are introduced early in the lifetime of a project, are
updated incrementally while driving the research, design and development
process, and require minimal overhead to maintain and keep up to date.

We focus in our presentation on software engineering requirements as are common
in research and development in academia and industry (which is sometimes called
research software engineering), thinking in particular about computational
science and data science. This is driven by our expertise and personal use
cases. The applicability of \nbval{} is wider though, and can similarly be
useful for Python-based software engineering outside a research context.

In section \ref{sec:chall-rese-softw} we derive and gather the requirements for
the design of the \nbval{} tool. Section~\ref{sec:org5b59194} describes the
practical use of \nbval. We close with a summary in section~\ref{sec:summary}.

\section{Challenges in research software engineering \& requirements for a
  notebook validation tool}
\label{sec:chall-rese-softw}

This section describes the main aspects and stages of a research
software engineering project and highlights associated challenges
which demonstrate the use cases and requirements for a notebook validation
tool. These requirements are summarised in section~\ref{sec:orgea58a1a}.

We note that these stages are not meant to provide a strict
categorisation nor to describe a linear project progression. They
should rather be seen as a rough mental model to help distinguish
different workflows and focus areas over a project's lifetime, and as
such they help highlight the different challenges associated with
each.

\subsection{Experimentation / prototyping}
\label{sec:experimentation_prototyping}

A research software engineering project may start with an
experimentation and prototyping stage (which may or may not use an
existing code base), based on a research question or hypothesis.  Often
the first results of such an initial computational or data science
study suggest new requirements for the next steps. For example, a
result from a simulation may indicate that the assumed physical model
is not accurate enough, so it needs to be changed.  The gained
insights lead to modifications of the underlying code and further
iterations.

Key requirements at this stage are: the ability for interactive
exploration; fast iterations with short feedback loops; and the
ability to document the process, as well as any interesting outcomes,
``on-the-fly'' with minimal overhead.

Jupyter notebooks are well-suited to support these requirements and
workflow. They provide the ability to combine descriptive text
(including \LaTeX-formatted equations) together with code segments and
their outputs in the same notebook document [12, 2]. The code segments
can be executed interactively, and the output from the execution is
inserted automatically into the document [12].
%


\newcommand{\ignore}[1]{}


Jupyter notebooks are a great tool for interactive explorations and
coding. However, it is not uncommon for the investigations and studies
over the course of a project to lead to the creation of many
notebooks, for example applying the same analysis to different data sets.


As the underlying code (which often lives in external modules or
packages outside the notebooks themselves) evolves, older notebooks can
become stale and stop working. As the project grows, it may become infeasible
to manually inspect and
re-execute all notebooks to ensure they are still working and produce
the same results as before.

Consequently, older notebooks often contain outdated or broken
code. At best, it is time-consuming to update these notebooks when
they need to be revisited later (e.g. to produce a plot for
publication). At worst, it can be impossible to fix the code in a
notebook if the underlying software has changed too much in the
meantime, potentially leading to the loss of results.

What is needed in these situations is a tool that can alert the developer of any
notebooks which contain broken code or produce different results when
re-executed. Ideally, the alert is raised as soon as possible after the change
has occurred~\cite{beck-tdd}.

\subsection{Maturing and stabilising; finding the right code design}
\label{sec:matur-stab-find}

When the approach and algorithms start to stabilise, it is important that the
code can be safely refactored~\cite{refactoring}, without accidentally breaking
or changing any existing behaviour and results. In order to be able to do this safely it is
essential~\cite{working-effectively-legacy-code} to have a suite of
tests in place which specify the desired behaviour of the code and of
the outputs it produces (e.g., ``system tests'' or ``acceptance
tests''). Such tests allow the developer to safely perform modifications on the code base
so that the new functionality can be implemented without altering or
invalidating any of the existing outputs. These tests should be run
frequently during the refactoring so that any bugs can be detected
quickly.

%

We will discuss testing further in the next section \ref{sec:testing}. Here we
note that in order to safely modify the code base it is crucial to have tests
available.

The Jupyter notebooks which were produced during prototyping and subsequent
explorations (see previous section \ref{sec:experimentation_prototyping})
provide a natural suite of system/acceptance tests for the underlying code base.

The requirement for \nbval{} is the ability to automatically re-execute an
existing notebook with a modified version of the underlying codebase and verify
that nothing is broken and that the outputs produced are still the same as
before. 

\subsection{Testing}
\label{sec:testing}

We discuss two aspects of testing scientific software.

Firstly, it can be challenging to write
meaningful formal tests. Frequently, an experienced researcher can
assess and verify a result reasonably quickly by inspection, but it
can be difficult and time-consuming to write a
formal test for it. (Examples: looking at a line plot vs. testing for
properties in the underlying sequence of values; assessing a vector
field plot vs. checking properties of this vector field formally; or
checking an array of values in a pandas dataframe.)

\begin{figure}[tb!]
  \centering \includegraphics[width=0.9\textwidth]{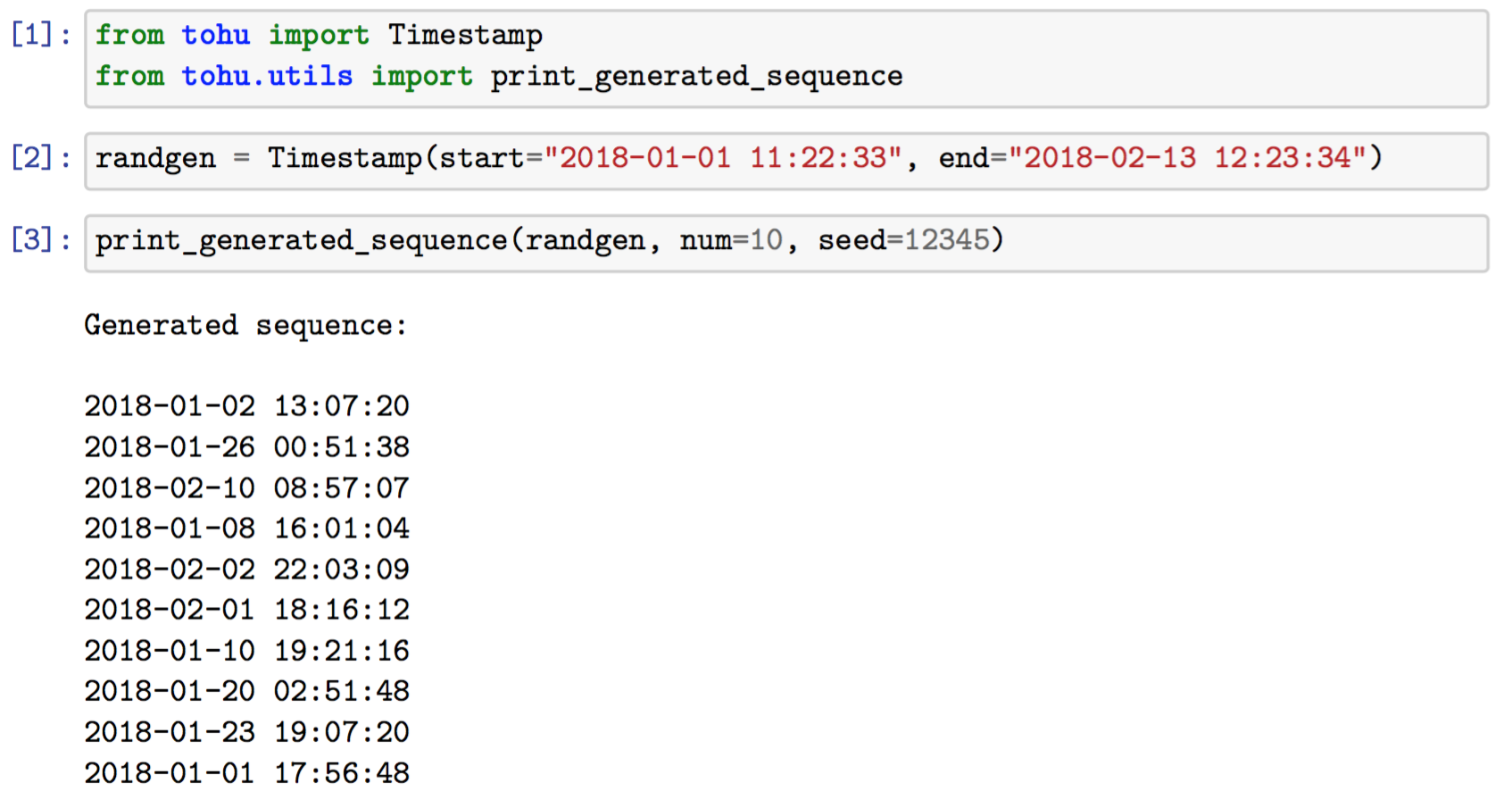}
  \caption{Excerpt from a Jupyter notebook which contains unit tests
    for \texttt{tohu} (a tool for reproducible generation of random
    synthetic data~\cite{tohu2020}). This example demonstrates how an \nbval-tested
    Jupyter notebook can simultaneously serve as documentation for how
    to use the code, and at the same time serve as a unit test for the
    generated output (in the last cell). The command is generating a sequence of
    random dates and times between a given start and end time.
    This type of output is easy
    for a human to verify by inspection but inconvenient to copy and paste
    into a sequence of formal assert statements. In addition, if the
    output changes, it is trivial to update the Jupyter notebook-based
    test by re-executing the cells, whereas manual update of assert
    statements would be more laborious.}
\label{fig:example_testing_notebook}
\end{figure}

Fig.~\ref{fig:example_testing_notebook} shows an example section of such a
testing notebook where the result is easy to visually inspect and confirm to be
correct but tedious to copy and paste into a sequence of formal assert
statements.

The figure also demonstrates how the process of writing tests can change when using
notebooks and \nbval: in conventional dedicated test code, one could write a
line of testing code \texttt{assert sum(40, 2) == 42} inside a test function.
Using Jupyter notebooks as a tool to define tests, one can write \texttt{sum(40,
  2)} in the code cell, and confirm on execution that the correct result
(\texttt{42}) is displayed, and then save the notebook. Once the tests are
defined this way, the execution of \nbval{} will report a pass if the next
execution of \texttt{sum(40, 2)} still produces \texttt{42}, and a fail otherwise.

Secondly, not all researchers have training in software
development~\cite{Hettrick2014}, so they may not have been trained in writing
tests, and may stick to manual checks because the overhead of writing formal
tests may feel too much of a burden, especially if the design and interface is
still changing and they are expected to be updated frequently. This provides an
incentive to delay automatic testing until later in the project, or to never
complete this.


%
%

Proponents of test-driven development (TDD)~\cite{beck-tdd} advocate
for writing tests upfront, before the actual implementation. This has
multiple benefits: it gives a clear indication when the coding task
for a given iteration is completed (namely, when all the tests pass);
and it helps identify the interface requirements without getting
caught up in the weeds of the implementation.  On the other hand, the
TDD approach can be impractical during prototyping, when one is trying
to identify whether an approach is even feasible, or trying to find
\emph{any} implementation (which can then be discarded and re-implemented
using a TDD approach).

A useful compromise is to use ``explorative sprints'' where one focuses
purely on the exploratory aspects, and learns enough about the
approaches and potential to then tackle the actual
implementation. As soon as this sprint leads to an insight that is worth
preserving one should be able to convert the notebook into a set of tests or documentation
which helps freeze the desired behaviour
in place. (This provides the reassurance that further changes or
additions to the code will not break the current implementation.)

A useful analogue is that of a ``ratchet wheel'', where each new \nbval-tested
notebook acts as an additional ``latch'' to prevent it from accidentally
releasing and turning to undesired behaviour.


\Nbval{} and notebooks can help convert those interactive and implicitly
defined checks into more formal and automatically verifiable tests
with minimal overhead, so they can be run regularly and
automatically and ideally be included in continuous integration
runs.

\subsection{Producing (reproducible) research outputs}
\label{sec:producing_research_outputs}

When the software has sufficiently matured that it can be used to produce
outputs that are subsequently published in academic journals, and -- for example
-- are used to drive healthcare, engineering or economic decisions and policy,
it is important to ensure these outputs are \emph{reproducible}: is all the
required information provided with the publication/report so that another party
could repeat the study, and get to the same conclusions?

Reproducibility is an important topic in its own right, and increasingly
expected by funding bodies and publishers (for example~\cite{NatureEditorial2017}). The
Jupyter Notebook is a useful tool to support this~\cite{Jupyter}: by driving
computational science and data science from Jupyter notebooks, the notebooks
serve as a detailed and complete transcript of all steps required to produce
the results in the paper. The computational code used is typically kept in
external libraries that are only called from the notebooks. A number of authors
have started to publish (Github) repositories that complement publications, and
provide one notebook per figure in the publication: by executing the notebook,
the figure can be reproduced (for example~\cite{Albert2016}), and we endorse
this as good practice.

With research codes, the underlying code may develop further, and this would
break the reproducibility of the notebooks.
The \nbval{} requirement for this use case is to be able to automatically
re-execute notebooks and to confirm that they produce the same results as
stored, or alert to a problem by reporting a failure.
Software solutions to address the problem when the code base changes are outside
the scope of this paper; possibly solutions include versioning of the underlying
code and/or to preserve the software environments as was used at time of
publication.

%
%
%

\subsection{Documenting software}
\label{sec:documentation}

\subsubsection{Introduction}

In order to share research with the world, it needs to be documented. The
minimum documentation for research outputs is a complete description of all
required steps to produce the research outputs (see
section~\ref{sec:producing_research_outputs}).

Sometimes the research software that has been used for a study has value in its
own right: it may be re-usable for a follow up study, or even evolve into a
framework that allows simulating / analysing many other systems, or can be
developed further more effectively than recreating a new tool from scratch.

In that case, more documentation of the software is required:
tutorials, how-to guides, contextual explanations and references (including
application interface definitions)~\cite{procida-documentation}).

Research software is likely to keep changing over time: These frequent changes
can be a challenge for keeping documentation up-to-date, in particular if the
people working on the software are so familiar with it that they rarely need to
consult the documentation.

%

\subsubsection{Use of Jupyter notebooks for documentation}

The creation of software documentation using Jupyter notebooks can be very
effective, combining descriptive text, equations, and code in the same
notebook document~\cite{discretisedfield-docs,pandas-docs}. By executing the
code cells, the output from the execution is inserted automatically
into the document. For scientific purposes, figures and images
created from commands is a common use case, and these are also inserted into the
notebook~\cite{pandas-docs} as an output. If a figure or a code segment needs
updating, it is sufficient to change the code segment and to re-execute the cells. 

If there are any changes to the software that is used and described in the
notebook, the documentation can be updated by re-executing the notebook (as this
will update the text-based and media outputs that are generated in the
notebook). This partly automates updating of the documentation, and reduces
manual effort.

Because notebooks can be used as sources for Sphinx~\cite{sphinx,nbsphinx} and
MkDocs~\cite{mkdocs,mknotebooks} documentation, any computational and data
science data exploration can become a part of the documentation, and is often
useful as a how-to guide or tutorial.

This documentation embedded in notebooks needs to be contrasted to the more
conventional method of documentation generation, where descriptive text is stored
in one document (for example in the restructuredtext or markdown format), which
may include code segments that either need to be copied and pasted or included from
a separate file. Inclusion of figure files is realised through a command that
lists the name of the file containing the figure on disk, and inclusion of that
file at a later documentation compile time. In this case, updating the code
segment that leads to a figure requires manual transfer of the code snippet into
an executable file/environment, and execution of this, and taking care to make
sure the right figure file is updated as a result of this execution. Once this
is done, the whole documentation needs to be compiled to check that the
resulting html or pdf output is correct.

\subsubsection{Executable ``live'' documentation}
\label{sec:org04d0550}
We mention in passing the possibility of executing Jupyter notebooks in the
cloud using the Binder project and the mybinder.org instance~\cite{binder}:
Using this service, Jupyter notebook based documentation can be turned into
executable documentation (as long as the software and data required are
available on the Internet). The reader can jump into an executable notebook that
is hosted on the remote machine of the mybinder service, and execute and modify
examples from the documentation without having to install any software locally
(only a web browser is required)~\cite{discretisedfield-docs}. This reduces the
barrier towards learning a new piece of software. Binder is also useful for
reproducible science~\cite{reproducible-micromagnetics}.

\subsubsection{\Nbval{} for documentation}
Assuming the documentation is created using Jupyter notebooks, a tool such
as \nbval{} can address some of the the documentation challenges: \Nbval's role is to help
capturing anything that stops working, for example by using \nbval{} as part of
the continuous integration to check that the documentation notebooks execute and
produce the output as expected.

This is helpful if the documentation is created up-front or alongside the
development (in an iterative fashion) as it alerts the developers when new
interfaces or behaviour make the documentation notebooks incorrect. It is also
helpful when documentation is created at a particular point in the project's
lifecycle: it is typically not revisited by the authors (as they don't need to
consult the documentation) and without periodic testing of the documentation, it
may become inaccurate when further development is taking place.

Using \nbval{} to validate documentation as part of the continuous integration,
will also alert the developers to changes in behaviour of third party libraries
that are used in a given project.

\subsubsection{Use of documentation as tests}
\label{sec:org10779d9}
An existing Jupyter notebook that documents or showcases the behaviour
of a piece of software by showing code snippets together with the
output values that the code produces, can also be seen as a set of
software tests (see Fig.~\ref{fig:example_testing_notebook} for a simple
example). If this is a low level piece of code, we can interpret this
as a unit test, whereas if it tests features of a larger module or a
combination of modules, it can be seen as a system
test~\cite{van-vliet}.



%
%
%

\subsection{Requirements for \nbval}
\label{sec:orgea58a1a}

Drived by these observations, we desired a tool that helps to keep Jupyter notebook
based documents up-to-date, and that allows to read the combination of a code
cell and the stored output as a regression test: can that same output be recomputed from
the input? This should be done automatically, so that for each notebook cell, we
have a PASS or FAIL outcome. This can then be integrated into existing
unit test frameworks, and allow us to (i) use existing notebooks as automatic
tests, and (ii) to check if existing documentation notebooks are still
up-to-date. The tool NoteBook VALidate (\nbval) has been developed to fulfil these
requirements.

\section{The \nbval{} tool}
\label{sec:org5b59194}
We provide an overview of the \nbval{} design and implementation, usage options,
and comment on its limitations. The tool is available as open source~\cite{nbval-github}
and more details are provided in the documentation~\cite{nbval-documentation}.

\subsection{A plugin to \pytest}
\label{sec:orgbfdc0cf}
To re-use existing functionality as much as possible, we have developed the
current version of \nbval{} as a plugin to the \pytest{} tool~\cite{pytest}. \Pytest{}
can scan subdirectory trees for files that match particular patterns. With the
\nbval{} plugin activated, \pytest{} will find files with the extension
\texttt{.ipynb}, and validate each of these notebooks.

The Jupyter notebook format \texttt{.ipynb} stores outputs and inputs for
each cell.
Validating\footnote[1]{From a software engineering terminology perspective, \emph{verification} may
  have been a better term than \emph{validation}~\cite{van-vliet}.}
the notebook means to rerun the notebook and to make sure that it 
generates the same outputs as have been stored.
Fig.~\ref{fig:demo-each-cell-a-test} shows example output from validating a
notebook.

\begin{figure}[t!]
  \centering
\begin{lstlisting}[basicstyle=\ttfamily\scriptsize,frame=single]
$ py.test --nbval -v 03-data-types-structures.ipynb
============================= test session starts =========
platform darwin -- Python 3.8.0, pytest-5.3.1, ...
plugins: nbval-0.9.3
collected 137 items                                                            

03-data-types-structures::ipynb::Cell 0 PASSED       [  0%]
03-data-types-structures::ipynb::Cell 1 PASSED       [  1%]
03-data-types-structures::ipynb::Cell 2 PASSED       [  2%]
...
03-data-types-structures::ipynb::Cell 136 PASSED     [100%]

===================== 137 passed in 6.08s =================
\end{lstlisting}
  \caption[Each cell is a test.]{Example output from running \nbval{} to validate
    one Jupyter notebook: each code cell in the notebook becomes a test. The
    chosen example is chapter~3 of a book~\cite{fangohr-python-book} which is composed
    of a sequence of Jupyter notebooks (one per chapter), and this \nbval{} test is
    used as part of the continuous integration on Travis~CI. The \texttt{--nbval} flag
    requests that \pytest{} will process \texttt{ipynb} files. The
    \texttt{-v} requests more verbose output from \pytest{} than is
    standard.}
  \label{fig:demo-each-cell-a-test}
\end{figure}

\subsection{Usage}
\label{sec:orgbe5d756}

We use \nbval{} by running \pytest{} with the
\texttt{--nbval} or \texttt{--nbval-lax} flag (the difference is described below).
This causes \pytest{} to collect files with a \texttt{.ipynb} extension and
pass them to \nbval, in addition to finding and running more conventional Python
tests.

As a basic check, \nbval{} expects all cells to run without errors.
If executing a cell produces an error, it counts as a failure regardless of
any output it produces. If a cell runs without an error, there are several
ways to control whether its output should be checked.

\subsection{Controlling output checks}

Some code is meant to produce entirely consistent output given the same input.
However, in many cases some output is expected to vary --- e.g. if a timestamp is
displayed, or if the code includes any random elements. Various features of
\nbval{} allow it to check only selected parts of the output.

There are two possible ways to approach this. In lax or relaxed mode (with the
\texttt{--nbval-lax} option), no output is checked by default, but the author
of the notebook can mark cells as described below to have their output checked.
In strict mode (with the \texttt{--nbval} flag), all output is checked
except for cells marked to indicate otherwise.

Lax mode provides a fairly quick way to validate notebooks written primarily as
documentation, which will detect in many cases if code changes break the
examples.

Using strict mode, it generally takes some work to make a notebook pass,
as one needs to deal with any cells which may produce different output while
still working correctly. The strict mode will also pick up subtle changes in the
way output is produced. For example, if the text-based formatting of numpy arrays was to change by a space
somewhere, the output cells displaying numpy arrays would fail.
If a manual review of the failures shows that the change is not a problem,
rerunning the notebook and saving the new output will make these tests pass
in the future.

\subsection{Cell markers}

Code comments or cell tags can be used to enable or disable output checking
cell-by-cell. Cell tags are a Jupyter feature, where short strings can be stored
in the metadata of each cell. These are not visible by default, but a cell
toolbar can be enabled to see and edit them.

Adding a tag \texttt{nbval-check-output} to a cell tells \nbval{} to check its
output in relaxed mode, while \texttt{nbval-ignore-output} disables output
checking in strict mode. The same words can also be used as comments in the code,
but here they are uppercase and separated by underscores, e.g.:

\centerline{\includegraphics[width=0.7\textwidth]{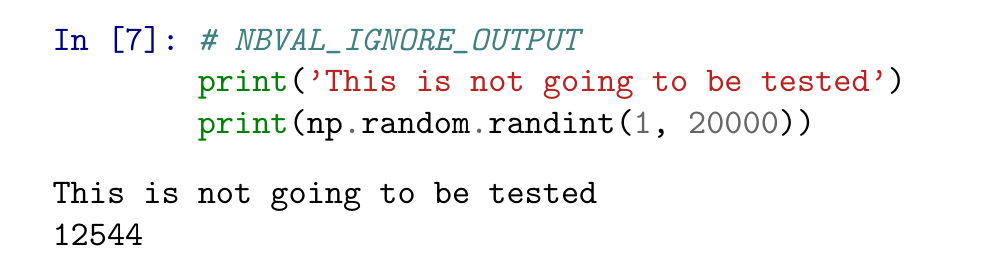}}

\nbval{} also recognises some other markers specified in the same ways:
\texttt{nbval-skip} will skip over a code cell entirely, and
\texttt{nbval-raises-exception} or \texttt{raises-exception} will ignore an
error from executing a cell, which would normally be shown as a failure.

\subsection{Example for successful (PASS) and unsuccesful validation (FAIL)}

A code cell containing deterministic code should pass. For example:
\begin{lstlisting}[basicstyle=\footnotesize\ttfamily]
print('Hello World')
\end{lstlisting}

A notebook containing one cell that will create different output when
executed again, will fail in strict mode. Figure~\ref{fig:fail-time} shows
output from a failing test for a code cell containing these lines:
\begin{lstlisting}[basicstyle=\footnotesize\ttfamily]
import time
time.asctime()\end{lstlisting}
\begin{figure}[tb!]
  \centering
\begin{lstlisting}[basicstyle=\ttfamily\scriptsize,frame=single]
(nbval) > $ py.test --nbval time.ipynb 
collected 1 item
time.ipynb F                                         [100%]
=================================== FAILURES ==============
______________________________ time.ipynb::Cell 0 _________
Cell 0: Cell outputs differ
Input:
time.asctime()

Traceback:
 mismatch 'text/plain'

 assert reference_output == test_output failed:
  "'Thu Dec 12 16:20:49 2019'" == "'Thu Dec 12 16:24:38 2019'"
  - 'Thu Dec 12 16:20:49 2019'
  ?                 ^ ^^
  + 'Thu Dec 12 16:24:38 2019'
  ?                 ^ ^^
========================= 1 failed in 0.72s ===============
\end{lstlisting}
\caption{An example for a mismatch in output, which is reported as a failure
  when running \nbval{} in strict mode (using \texttt{--nbval}):
  The timestamp saved is different from the timestamp when the cell was executed by \nbval.
  The ``\texttt{-}'' sign before ``Thu Dec 12 \ldots''
  indicates that this line of output is present in the saved output in file
  \texttt{time.ipynb} but missing in the re-computed output, and
  the \label{fig:fail-time} ``\texttt{+}'' sign two lines below indicates that
  this line is new in the computed output and was not present in the saved file.
}
\end{figure}

\subsection{Sanitizing output with regular expressions}\label{regex-output-sanitizing}

\nbval{} supports more fine-grained handling of variable outputs,
by matching and sanitising patterns such as timestamps or memory addresses
within output that it is comparing.
The user can specify a sanitizing file at the command prompt
using the following flag:

\begin{lstlisting}[basicstyle=\footnotesize\ttfamily]
$ py.test --nbval x.ipynb --sanitize-with my_sanitize_file
\end{lstlisting}

This sanitize file \texttt{my\_sanitize\_file} contains a number of regular expressions and replacement strings.
It is recommended to replace the removed output with a recognisable marker;
for instance a timestamp like \texttt{16:44:06} might be replaced with \texttt{TIMESTAMP}.
This replacement will be done in both the saved and the recomputed output
before comparing them, so differences only in the replaced text do not cause
failures.
The marker is useful to understand failing tests.

\subsection{Figures and multimedia elements}
By default, only text output is compared~\cite{nbval-documentation},
but \nbval{} can integrate with \texttt{nbdime}~\cite{nbdime} to display rich comparisons
when outputs differ. If this option (\texttt{--nbdime}) is used, image outputs
(PNG and JPEG formats) are also compared.

\section{Summary}
\label{sec:summary}

\Nbval{} is a plugin to \pytest, which allows to check that the output saved in
the past in a notebook file is consistent with output computed today. Use cases
include reproducible science, checks that deployed software behaves as its
documentation suggests, that tutorials, manuals and textbooks remain up-to-date,
and to help notice if support libraries change their behaviour. Deployment of
\nbval{} in continuous integration automates this process, and notebooks can
serve as additional system tests and provide additional test coverage. We note
that unit, integration and system tests can be written using Jupyter notebooks;
which reduces the effort of formulating the test statement. Notebooks in
combination with \nbval{} and continuous integration can be used to develop
documentation and tests as part of the design exploration and implementation
phase in an agile manner.

Conceptionally, \nbval{} provides testing functionality for notebooks which is similar
to doctest for Python documentation strings. Due to the nature of notebooks, these can be
used more flexibly than docstrings as outlined above.

\subsubsection{Funding acknowledegments}
We acknowledge support from the Horizon 2020 European Research Infrastructure
project OpenDreamKit (676541) and Photon and Neutron Open Science Cloud (PaNOSC)
project (823852), EPSRC's Centre for Doctoral Training in Next Generation
Computational Modelling \linebreak (EP/L015382/1), EPSRC's Doctoral Training Centre in
Complex System Simulation (EP/G03690X/1), CONICYT Chilean scholarship programme
Becas Chile (72140061), the Gordon and Betty Moore Foundation through Grant GBMF
\#4856, the Alfred P. Sloan Foundation and by the Helmsley Trust, the University
of Southampton, Simula, and European XFEL.
%
%

\end{document}